\documentclass[a4paper, prd, twocolumn, showpacs]{revtex4}
\usepackage{float}
\restylefloat{table}

\usepackage{amssymb}
\usepackage{amsmath}
\usepackage{color, mathrsfs}
\usepackage{graphics, graphicx, amsmath, amsthm, amssymb}
\usepackage{bbold}

\usepackage[normalem]{ulem}
\newcommand\redout{\bgroup\markoverwith
{\textcolor{red}{\rule[.5ex]{2pt}{0.4pt}}}\ULon}
\usepackage{xcolor}

\usepackage{verbatim} 

\newcommand{\mr}{\mathrm}

\newcommand{\bb}{\mathbb}
\newcommand{\ms}{\mathscr}
\newcommand{\frk}{\mathfrak}

\usepackage{psfrag}
\usepackage{mathcomp}
\usepackage{subfigure}
\newtheorem{Def}{Definition}
\newtheorem*{Pos}{Postulate}
\newtheorem{Thm}{Theorem}
\newtheorem{Pred}{Prediction}
\newtheorem*{Fact}{Facts}

\usepackage{color}
\usepackage{soul}

\begin{document}

\author{Guowu Meng}

\affiliation{Department of Mathematics, HKUST, Clearwater Bay, Kowloon, Hong Kong} 
\date{June 18, 2012}

\title{What are the Fundamental Matter Particles?}

\begin{abstract}
Quantum theory and Lorentz structure are the twin pillars of fundamental physics today. With quantum theory kept and Lorentz structure replaced by Euclidean Jordan algebra --- a more fundamental structure, one naturally arrives at the notion of abstract fundamental matter particles. 

These abstract particles fall into distinct abstract universes according to their symmetry groups. If it is assumed that the charged particle count for such an abstract universe is $32$, then this abstract universe must be conformally-symmetric and $11$-dimensional Lorentzian when it is extremely hot; furthermore, it matches our real world universe in quite a few aspects, from the charged particle content to the existence of dark matter. Based on this match, a few predictions can be made: 1) the electric-weak force symmetry must be broken if the macroscopic spatial dimension is 2 or higher, 2) there are infinitely many generations of quarks and leptons, and 3) there exists a 5th fundamental force.
 
This 5th force predicts quark mixing and the related CP violation because it transforms quarks among its various generations and violates the CP symmetry. A quantitative check concerning the Cabibbo-Kobayashi-Maskawa matrix entries shows a good agreement between experiments and the 5th force based computations.

 \end{abstract}
\maketitle

\section{INTRODUCTION}

Despite the great success of the Standard Model of particle physics, our current understanding of physics is still rather incomplete at the fundamental level. For example,
quarks and leptons must be introduced into the theory manually, and their charges and spin cannot be determined through the current theory, and instead, must be determined by experiments. 

This situation is somewhat similar to that of quantum theory before Lorentz structure was introduced. During that time, electron spin was also unpredicted by the theory, and thus was added in manually. However, when quantum theory was combined with Lorentz structure, electron spin no longer needed to be manually introduced, and moreover, antimatter was predicted.

\subsection{The Twin Pillars}
It is clear that the twin pillars of fundamental physics today are \emph{quantum theory} and \emph{Lorentz structure}, and they descended from the twin pillars in Newton's time via the two great physics revolutions in the early 20th centrury. Schematically this can be summarized as follows: \begin{center}
 \redout{\em Classical mechanics} + {\em Galilean structure}
$$\downarrow\begin{matrix} \mbox{quantum}\\ \mbox{modification}\end{matrix}$$
{\em Quantum theory} +  \redout{\em Galilean structure}
$$\downarrow\begin{matrix} \mbox{relativity}\\ \mbox{modification}\end{matrix}$$
{\em Quantum theory} + {\em Lorentz structure}.
\end{center}
Given the fact that the Standard Model still has many places where experimental findings are manually incorporated, most physicists would agree that these twin pillars of fundamental physics still need further improvement. In this article, the following further modification:
\begin{center}
{\em Quantum theory} +  \redout{\em Lorentz structure}
$$\downarrow\begin{matrix} \mbox{new}\\ \mbox{modification}\end{matrix}$$
{\em Quantum theory} + {\em Euclidean Jordan algebra}
\end{center}
is proposed. This modification is comparatively conservative because Lorentz structure is a secondary structure hidden inside Euclidean Jordan algebra, rather than an approximation to it \footnote{Therefore, unlike the quantum and the relativity modifications, there is no new fundamental constant involved in this modification.}. In fact it might have been contemplated by physicists who are familiar with Euclidean Jordan algebra. It is not a surprise that this contemplation fails to appear until now because the key ingredient associated with Euclidean Jordan algebras, i.e.,  generalized Kepler systems, were discovered only recently.

With this modification of the twin pillars, quarks and leptons, dark matter, the four fundamental forces and the broken electric-weak symmetry, matter generations, and other experimentally found phenomena, appear naturally in the theoretical framework. Moreover, a 5th fundamental force is predicted.

\subsection{The Fifth Force}
This 5th force transforms quarks among its various generations and violates the CP symmetry, so it immediately predicts quark mixing and the related CP violation, which in fact was a phenomena observed in labs by J. Cronin and V. Fitch about 50 years ago. Thus, there are two theories as of now, the one based on this 5th force which predicts this phenomena and the established Cabibbo-Kobayashi-Maskawa (CKM) theory \cite{Cabibbo63} which was invented about 40 years ago to explain this phenomena. 

In the CKM theory, a key assumption is that the electric-weak symmetry mixes quarks (but not leptons) from different generations, something mysterious and not supported in the theoretical framework presented in this article. This is a serious problem --- at least one of these two theories must be wrong. By testing them against the recent experimental data, it is found that the CKM theory fails the test, whereas the one based on this 5th force withstands the test well, in both accuracy and precision. For example, for the CKM matrix entry $V_{cd}$, we have
\begin{table}[H]
    \begin{tabular}{|c|c|c|c|}
        \hline
       & {\bf experiment} & $\begin{matrix}\mbox{CKM theory based}\\ \mbox{computation}\end{matrix}$  & $\begin{matrix}\mbox{5th force based}\\ \mbox{computation}\end{matrix}$\\ \hline
               $|V_{cd}|$  & $\mathbf {0.230\pm 0.011}$ & $0.46\pm 0.48$ & $0.226\pm 0.034$\\ 
        \hline
    \end{tabular}
\end{table}
\vskip -15pt
\noindent For more details, please consult Ref. \cite{mengmay12}.

\subsection{Outline}
This article will include ten sections. In section \ref{S: Kepler}, the generalized Kepler systems \cite{meng}, which owe their existence to Euclidean Jordan algebra, are reviewed. In section \ref{S: Definitions}, the notion of abstract particles and abstract physical universes are introduced. In section \ref{S: Postulate}, a postulate about our real world universe is made. This postulate singles out a unique abstract physical universe to match with our real word universe, and the details is presented in section \ref{S: Initial Matching}. Some further matchings are made in section \ref{S: Further Matching}. Section \ref{S: visible states} contains quite a lot of technical details, but mathematically it makes a few concepts such as broken symmetry, exact symmetry, and quark and lepton generations more precise. Based on this technical analysis, a 5th fundamental force is predicted in section \ref{S: 5th force}, in which indirect experimental evidence for this 5th force is also presented. In section \ref{S: loose ends}, some loose ends are pointed out offering directions for further research. The article ends with section \ref{S: conclusion}, which summarize the major results achieved in this article.

Note that in this article, any Lie groups we shall mention or use is assumed to be connected, and subgroup relations are often up to coverings. Readers who do not like Lie groups can just pretend that we are dealing with Lie algebras and Lie subalgebras, with essentially nothing lost.

\section{Generalized quantum Kepler Systems}\label{S: Kepler}
A key concept of this article is the notion of abstract fundamental matter particles, which shall be defined as generalized quantum Kepler systems, i.e., the well-known quantum Kepler system and its vast generalization \cite{meng} discovered in recent years.  

These generalized Kepler systems, which are all super integrable models resembling the Kepler system, are discovered by this author to be naturally associated with simple Euclidean Jordan algebras. Therefore, with Lorentz structure replaced by Euclidean Jordan algebra and quantum theory kept, (abstract) fundamental matter particles naturally appear.

\subsection{Fundamental Objects in Physics}
To motivate the definition of abstract fundamental matter particles, it is worth to make a small digression here to re-examine the fundamental objects in some great past periods of physics.

In Newton's time, the fundamental object in physics was the solar system. In relation to this, a math-minded person would possibly wonder how the simplest solar system can be described abstractly. A good abstraction for the simplest solar system is the classical Kepler system. Also, in the beginning of the 20th century, one of the fundamental objects in physics was the atom. In this case, a math-minded person would also wonder how the simplest atom can be described abstractly. An equally good abstraction for the simplest atom would be the quantum Kepler system. Similarly, ever since the 1960s, a fundamental object in physics has been the fundamental matter particles --- quarks and leptons. However, if a math-minded person would wonder in this case as to how quarks and leptons can be described abstractly, there is no obvious answer. But if we believe great ideas can be used repeatedly in mathematics and physics, it is tempting to say that the answer has to do with the Kepler system again. Indeed, that is the case, but in a more sophisticated way.  

\subsection{Kepler System and its Family}
Recall that, the {\it quantum Kepler system} is the quantum system with hamiltonian $H=-{1\over2}\Delta- {1\over r}$ and wave functions being square-integrable functions on $\bb R^3\setminus\{(0,0,0)\}$. Here $\Delta$ is the Laplace operator, $r$ is the distance from the origin, and all parameters have been set to be $1$. 
The followings are some basic facts about this wonderful quantum system: 
\begin{itemize}
\item For $I=0$, $1$, $2$, $\ldots$, the $I$-th energy eigenvalue is $E_I=-{1/2\over (I+1)^2}$ and the $I$-th energy eigenspace $\ms H_I$ is the highest weight representation of $\mr{SO}(4)$ with highest weight $(I, 0)$. 

\item
$\bigoplus_{I\ge 0} \ms H_I$   
(more precisely its $L^2$ completion) is the unitary lowest weight representation of $\mr{SO}_0(4,2)$ (the identity component of $\mr{SO}(4,2)$) with lowest weight $(0, 0, 1)$.  
\end{itemize}

The Kepler system is a single member of a small family indexed by a discrete parameter $\mu\in {1\over 2}\bb Z$. The member with index $\mu$ is the quantum system for which the wave functions are square-integrable sections of the canonical line bundle over $\bb R^3\setminus\{(0,0,0)\}$ with the first Chern number $2\mu$ and canonical connection $A$, and its hamiltonian is
$$
H_\mu = -{1\over 2}\Delta_A+{\mu^2\over 2r^2}-{1\over r}
$$
where $\Delta_A$ is the Laplace operator twisted by the bundle.  The basic facts about the Kepler system generalize as follows: 
\begin{itemize}
\item For $I=0$, $1$, $2$, $\ldots$, the $I$-th energy eigenvalue is $E_I=-{1/2\over (I+1+|\mu|)^2}$ and the $I$-th energy eigenspace $\ms H_I$ is the highest weight representation of $\mr{Spin}(4)$ with highest weight $(I+|\mu|, \mu)$.

\item 
$\bigoplus_{I\ge 0} \ms H_I$   
is the unitary lowest weight representation of $\widetilde{\mr{SO}}(4,2)$ with lowest weight $(-\mu, -|\mu|, 1+~|\mu|).$
\end{itemize}

Here,  $\widetilde{\mr{SO}}(4,2)$ denotes the universal cover of the identity component of $\mr{SO}(4,2)$ and is referred to as the \emph{dynamic symmetry group} in the literature. Since all these quantum systems are defined on $\bb R^3\setminus\{(0,0,0)\}$, each of these infinite dimensional representations has the \emph{GK dimension} \footnote{ If $V$ is the space of polynomials in $n$ variables with the natural degree grading, then its GK dimension is $n$. Generalizing this, if $V=\bigoplus_{I\ge 0} V_I$ is a graded vector space with finite dimensional graded components, its GK dimension is defined to be 
$\lim_{k\to \infty}{\ln (\sum_{I=0}^k \dim V_I)\over \ln k}$. } equal to $3$.  

This small discrete family of quantum systems of GK dimension $3$ plus another small continuous family of quantum systems of GK dimension $4$ exhaust all generalized Kepler systems with dynamic symmetry group $\widetilde{\mr{SO}}(4,2)$. Here, the family of quantum systems of GK dimension $4$ is indexed by a continuous parameter $\lambda > 1$, and for a member with index $\lambda$, its the configuration space is the 4D space bounded by the future light cone in Minkowski space, and its Hilbert space of bound states $\bigoplus_{I\ge 0} \ms H_I$ is the unitary lowest weight representation of $\widetilde{\mr{SO}}(4,2)$ with lowest weight $(0, 0, \lambda)$. 

The preceding paragraphs gives a sketch of one big family of generalized quantum Kepler systems. Its associated simple euclidean Jordan algebra is the one hidden behind the Minkowski space. One can put the ``stamp" $\widetilde{\mr{SO}}(4,2)$ on this big family. In the mathematical abstraction given later, these generalized quantum Kepler systems shall be called abstract particles, and they form the abstract universe $\widetilde{\mr{SO}}(4,2)$, though an uninteresting one. 

\subsection{Generalizations}
There are many other big families of generalized quantum Kepler systems, stamped by
some simply-connected real non-compact Lie groups. These Lie groups form the following list: 
\begin{eqnarray}
\mr{SL}(2, \bb R), \quad \mr{SO}(2, n+1) \; (n\ge 2),\quad \mr{E}_{7(-25)},\nonumber\\
\quad \mr{SL}(2n, \bb R), \quad \mr{SU}(n,n), \quad \mr{SO}^*(4n)\quad (n\ge 3)\nonumber
\end{eqnarray}
up to covering and connectedness. 

Without going to too much details,  it is worth to mention a common feature of generalized quantum Kepler systems: the Hilbert space of bound states of a generalized Kepler system always forms a unitary lowest weight representation \footnote{The unitary lowest weight representations for non-compact real simple Lie groups (must be of hermitian type) have been completely classified in Ref. \cite{EHW82}. However, some of them, for example, the trivial representation and vector representations with the maximal possible GK-dimension, cannot be realized by generalized Kepler systems, although in general, representations of either minimal positive GK-dimension or scalar type (except the trivial one) can always be realized by a unique generalized Kepler system. } for a unique simply-connected real non-compact Lie group in the above list. It is also worth to mention that, to understand a generalized Kepler system, one must start with a simple Euclidean Jordan algebra, and above list of Lie groups are precisely the list of conformal groups of simple Euclidean Jordan algebras. For more details on generalized quantum Kepler systems, which is not absolutely needed here, please consult Ref. \cite{meng}.

\section{Major Definitions}\label{S: Definitions}
Let us start with a wonderful idea of E. Wigner \cite{Wigner1939}: an elementary particle (in the sense of Wigner) is a quantum system such that its different quantum states give rise to a unitary irreducible representation of the Poincar\'{e} group. From the mathematical point of view, Poincar\'{e} group is not a nice group, indeed, many unwanted particles appear in the Wigner's classification list of elementary particles. Also, there is no natural quantum system attached to a Wigner elementary particle. With this in mind, let us make the following key definition.
\begin{Def}(Abstract Particle)
A quantum generalized Kepler system is called an {\bf abstract particle} and its bound states are called the {\bf quantum states} of that abstract particle.  
\end{Def} 
For an abstract particle, $V$ is used to denote either its set of quantum states or the particle itself, and $V_0$ is used to denote its set of ground states. The dimension of $V_0$ is always finite and shall be referred to as the {\bf particle count} of that abstract particle $V$. Depending on whether this particle count is $1$ or not, the particle is said to be of {\bf scalar type} (i.e., a dark matter particle) or of {\bf vector type} (i.e., a charged matter particle). The GK-dimension of an abstract particle is simply the GK-dimension of the vector space of its quantum states, graded by the energy level of the corresponding generalized Kepler system.

Since the abstract particles fall into various classes according to their dynamic symmetry groups, the following definition can be made. 
\begin{Def}(Abstract Universes)
An {\bf abstract universe} is the set of all (types of) abstract particles with a fixed dynamic symmetry group.
 \end{Def} 
The particles in an abstract universe shall be referred to as {\bf allowable particles} in the abstract universe. The fixed dynamic symmetry group shall hereon be referred to as the {\bf abstract universe symmetry group} and in fact determines this abstract universe, so we shall use $G$ to denote either an abstract universe or its symmetry group. 

The {\bf charged matter particle count} is defined as the sum of the particle count over the set of all types of allowable charged matter particles, and the {\bf dark matter particle count} is defined as the sum of the particle count over the set of all types of allowable dark matter particles. Note that for an abstract universe, while the charged matter particle count might be finite, the dark matter particle count is always infinity. 

Our real world universe has existed in multiple stages. Immediately after its creation, our universe was in an \emph{extremely hot stage}, in which it is expected that for every type of fundamental matter particle, there was an equal probability for each of its quantum states to appear. Currently, our universe is in a \emph{fairly cold stage}, in which it is expected that certain quantum states of the fundamental matter particles appear with probability essentially zero. These states shall be referred to as \emph{hidden states} and the others shall be referred to as \emph{visible states}. With this in mind, we make the next definition. 
\begin{Def}(Abstract Physical Universe)
An {\bf abstract physical universe} is an abstract universe $G$ such that, 1) only allowable particles can appear,
2) the ground states of each allowable particles are visible states, 3) there is a fairly cold stage, in which $G$ has a Lie subgroup $G_{exact}=G_c\times G_n$ (up to covering) with $G_c$ compact and $G_n$ non-compact, such that, the action of $G$, when restricted to $G_{exact}$, leaves invariant the set of visible states of each allowable particle.
\end{Def} 
For convenience, for a generic allowable particle $V$, we shall use  $V_{visible}$ to denote the set of its visible states. The 3rd condition in the above definition says that $G_{exact}$ acts on $V_{visible}$  unitarily for each allowable particle. Since $G_{exact}$ is a symmetry group of all $V$ s, $G_{exact}$ is called the {\bf exact} symmetry group of the abstract physical universe in that fairly cold stage.  

It may well be possible that all $V_{visible}$ s share a symmetry group of the form $G_c'\times G_n'$, but this symmetry group is not a Lie subgroup of $G$ up to covering. If that happens, we say that $G_c'\times G_n'$ is a {\bf non-exact} symmetry group of the abstract physical universe in that fairly cold stage, furthermore, $G_c'$ is {\bf broken} if $G_n'\subseteq G_n$ and $G_c'\supsetneq G_c$,  and $G_n'$ is {\bf broken} if $G_c'\subseteq G_c$ and $G_n'\supsetneq G_n$. Obviously both $G_c$ and $G_n$ are not broken (or {\bf un-broken}), and are called respectively {\it the unbroken gauge group} and {\it the unbroken (macroscopic) space-time symmetry group} of the abstract physical universe in that fairly cold stage. Here is a word of warning: even if $G_c'$ is a broken gauge group, it is still possible to find a non-compact Lie group $G_n'$ such that $G_c'\times G_n'$ is a Lie subgroup of $G$ up to covering; on the other hand, even if $G_c'\supsetneq G_c$, $G_c'$ is not an broken gauge group unless there is non-compact Lie group $G_n'\subseteq G_n$ such that $G_c'\times G_n'$ is not a Lie subgroup of $G$ up to covering.

All of these notions naturally extend to the very early stage of the physical universe, here we take $G_c$ to be trivial and $G_n=G$.

There is a mass function $m$: $V_{visible}\to [0, \infty)$ for each allowable particle which is invariant under an exact symmetry transformation, but not invariant under a non-exact symmetry transformation.

Recall that $V$ denotes an abstract particle and $V_0$ denotes its set of ground states. Let $K$ be the maximal compact subgroup \footnote{We take $K=\mr{Spin}(11)$ (not the usual $\mr{Spin}(11)\times \bb R$) for $G=\widetilde{\mr {SO}}(11, 2)$.} of $G$ and 
$K_{exact}:=K\cap G_{exact}$. Then $V_0$ forms a unitary representation for both $K$ and $K_{exact}$. These representations respectively are referred to as the {\bf particle contents} of $V$ when the abstract physical universe is in extremely hot stage or fairly cold stage respectively. For a given abstract particle in an abstract universe, while the particle count is the same in all stages, the particle content can vary. 

A major piece of information that this article is trying to convey is that there is a unique abstract physical universe (i.e., the photo) that matches our real world universe (i.e., the fugitive) in quite a few fronts, from the particle content to the Lorentz symmetry. To make sense of this statement, a postulate about our current universe must be introduced.

\section{A Postulate}\label{S: Postulate}
Surprisingly, to find a unique matching abstract universe, all is needed is
 a single postulate about our current real world universe. Recall that a fundamental matter particle is called {\bf dark}  if its charges (including spin) are all trivial, and is called {\bf charged} otherwise. 
 \begin{Pos}
There are exactly $32$ types of fundamental charged matter particles in our current real world universe.
\end{Pos}
In our current real world universe the fundamental charged matter particles are quarks and leptons and there are at least three generations of them, and they all have spin $1/2$. Since there is no content difference among various generations of quarks and leptons, as far as particle count is concerned, there are 32 or 30 of them, depending on whether the right-handed neutrino (and its antiparticle) exists or not. These two particles, if they exist,  can only participate in the gravitational interaction, and that make them very hard to get detected due to the extreme weakness of gravity force.  So it is not unreasonable to assume their existence. Since not a single new charged particle as predicted in some theories has been found so far, this postulate looks quite reasonable. In any case, $32$ is much more beautiful, as shown by J. Baez and J. Huerta in Ref. \cite{Baez2010}. In that same reference, it is also explicitly shown that, with the right-handed neutrino (and its antiparticle) added to the classical Standard Model, the particle content $F$ under $G_{SM}:=\mr{U}(1)_Y\times SU(2)_I\times \mr{SU}(3)_c$ is the direct sum of
\begin{eqnarray}
\begin{array}{ll}
\mbox{left-handed leptons}&  {\bb C}_{-1}\otimes{\bb C}^2\otimes {\bb C}\cr
\mbox{left-handed quarks} &  {\bb C}_{1\over 3}\otimes{\bb C}^2\otimes {\bb C}^3\cr
\mbox{right-handed neutrinos}&  {\bb C}_0\otimes{\bb C}\otimes {\bb C}\cr
\mbox{right-handed electrons}&  {\bb C}_{-2}\otimes{\bb C}\otimes {\bb C}\cr
\mbox{right-handed up quarks}&  {\bb C}_{4\over 3}\otimes{\bb C}\otimes {\bb C}^3\cr
\mbox{right-handed down quarks}&  {\bb C}_{-{2\over 3}}\otimes{\bb C}\otimes {\bb C}^3,  
\end{array}\nonumber
\end{eqnarray}
and that the anti-particle content under $G_{SM}$ is $F^*$, i.e., the complex conjugate of $F$. It is also demonstrated there that
the action of  $G_{SM}$ factors through  $\overline{G_{SM}} :=G_{SM}/\bb Z_6$, 
$$\overline{G_{SM}} = \mr{SU}(5)\cap \left(\mr{SO}(4) \times \mr{SO}(6)\right)\subset \mr{SO}(10)$$
(since $\mr{SU}(5)$ is simply connected, $\mr{SU}(5)$, hence $\overline{G_{SM}}$,  can be and will be assumed to be a Lie subgroup of $\mr{Spin}(10)$.) and the spin representation of $\mr{Spin}(10)$, when restricted to $\overline{G_{SM}}$, is exactly the $G_{SM}$-module $F\oplus F^*$.

It should be pointed out that, what the nature really tells us is that $F\oplus F^*$ is a module of $G_{SM}\times \mr{Spin}(2)$ (not $G_{SM}\times \mr{Spin}(3)$), moreover, this module can be promoted to the positive spin module $S_+$ of $\mr{Spin}(12)$. That is because $G_{SM}\times \mr{Spin}(2)$ is a Lie subgroup of $\mr{Spin}(12)$ (up to covering) such that
$$
S_+\mid _{G_{SM}\times \mr{Spin}(2)} = F\oplus F^*.
$$

For the convenience of later analysis, some further simple facts from Ref. \cite{Baez2010} are listed below. First of all, as a subgroup of $\mr{U}(1)_Y\times \mr{SU}(2)_I$, we have
$$
\mr{U}(1)_e=\{(e^{i{\alpha\over 2}\hat Y}, e^{i\alpha \hat I_3})\mid \alpha \in \bb R\}. 
$$
Here, $\hat I_3$ is the element of $i\frk{su}(2)$ which becomes
$$
\left(
\begin{matrix}
1/2 & 0 \cr
0&  -1/2
\end{matrix}
\right)
$$
in the defining representation, and  $\hat Y$ is an element of $i\frk{u}(1)$.

Secondly, up to covering, as a Lie subgroup of $\mr{SU}(4)$,
$$
\mr{U}(1)_e\times \mr{SU}(3)_c=\left\{\left(\begin{matrix}e^{i3\alpha \hat Y} & 0\cr 0 & e^{-i\alpha \hat Y}g\end{matrix}\right)\mid g\in \mr{SU}(3), \;  \alpha \in \bb R\right\},
$$
where $\hat Y$ is an element of $i\frk{u}(1)$ which becomes $1\over 3$ in the defining representation of $\frk{u}(1)$. In fact,  up to covering, as a Lie subgroup of $\mr{SO}(8)$,
$$
\mr{U}(1)_e\times \mr{SU}(3)_c=\mr{SU}(4)\cap \left(\mr{SO}(2)\times \mr{SO}(6)\right)
$$
and can be lifted to a Lie subgroup of $\mr{Spin}(8)$. Therefore,  the defining representation of $\mr{SO}(8)$, when complexified and then restricted to $\mr{U}(1)_e\times \mr{SU}(3)_c$, becomes
$$
{\bb C}_1\otimes {\bb C}\oplus {\bb C}_{-{1\over 3}}\otimes {\bb C}^3\oplus C.C.,
$$
hence has no trivial component, a fact which implies that $\mr{U}(1)_e\times \mr{SU}(3)_c$ cannot be a Lie subgroup of $\mr{Spin}(7)$ (up to covering). In particular, $G_{SM}\times \mr{Spin}(2)$ cannot be  a Lie subgroup of $\mr{Spin}(11)$  (up to covering).

\section{Initial Matching with the real world universe}\label{S: Initial Matching}
With the postulate assumed, we are now ready to carry out the detective's job: matching the real world universe (the fugitive) with an abstract physical universe (i.e., a photo). We have substantial amount of knowledges about the real world universe, but we have to distinguish between hard facts and opinions. Hard facts cannot be disputed, but opinions are subject to revision. As an example, the existence of 30 observed (types of) quarks and leptons is a hard fact, but the existence of the right-handed neutrino and its antiparticle is just an opinion. The existence of many more types of new charged particles as predicted in some theories is an opinion, too.

In the process of matching, when we find a hard fact about the real world universe mismatched by an abstract physical universe $G$,
then we exclude $G$ as a possible candidate. However, if there is a mismatch of $G$ with an opinion about our universe, we cannot exclude $G$. It is entirely possible that all abstract physical universes are excluded in the end, if that happens, we have to accept it whether we like it or not. On the other extreme, there might be more than one equally plausible candidates. Fortunately that is not the case. 
\begin{Thm}\label{Thm1}
The abstract universe $G=\widetilde{\mr{SO}}(11, 2)$ is the only one with charged matter particle count equal to $32$, so it is the only possible abstract universe that stands a chance to match our real world universe.
\end{Thm}
From hereon, we shall make predications based on the assumption that the abstract universe $G=\widetilde{\mr{SO}}(11, 2)$ is indeed a mathematical photo of our real world universe.

It is expected that the real world universe is more symmetric when it is hot and less symmetric when it is cold, and has the largest possible symmetry when the temperature $T\to \infty$. Therefore, we have our first prediction.

\begin{Pred} The real world universe must be conformally symmetric $11$-dimensional Lorentzian when it is extremely hot. 
\end{Pred}
\noindent Note that this is also a  prediction from $M$-theory \cite{Mtheory}.  

We assume hereon that $G=\widetilde{\mr {SO}}(11, 2)$. In this abstract universe $G$, Ref. \cite{meng} tells us that, there is one type of charged matter particle with particle count $32$ and GK dimension $10$ --- the spatial dimension of the early universe, one type of dark matter particle with GK dimension $10$ (referred to as type I dark matter), uncountably many type of dark matter particles with GK dimension $11$ (referred to as type II dark matter) --- the spacetime dimension of the universe at birth, and no other type of fundamental matter particles; moreover,  the lowest weights of the corresponding representations are $(-{1\over 2}, -{1\over 2}, -{1\over 2}, -{1\over 2}, -{1\over 2},  5)$,  $(0, 0, 0, 0, 0,  {9\over 2})$, and
$$
(0, 0, 0, 0, 0,  \lambda)\quad \lambda > {9\over 2}
$$
respectively. Here comes the second prediction.

\begin{Pred}
The real world universe can have dark matter particles, i..e, particles carrying trivial charges and spin. Moreover, the charged matter particle content is exactly $F\oplus F^*$ as described in section \ref{S: Postulate}.
\end{Pred}
Here is the reason for the 2nd statement in this prediction: for the unique type of charged matter particle, $V_0$ is the spin representation of $\mr{Spin}(11)$, so $V_0=F\oplus F^*$ under $\mr{Spin}(10)$. 

Note that, the predicted charged particle content is a fact for the 30 particles and an opinion for the extra two particles in particle physics. The predicted existence of dark matter is an opinion in particle physics/cosmology, too. In our theory we may or may not include dark matter, similar to the situation in general relativity in which the cosmology constant may or may not be included in the Einstein equation. That is why we use the word ``can" rather than ``must" in the statement. Also, dark matter in our theory cannot interact through the weak interaction, and that is different from a current dominant opinion in the physics community.
   
\vskip 10pt
The proof of Theorem \ref{Thm1} is based on the knowledge about the generalized Kepler systems and goes as follows: Since the charged matter particle count is finite, there must be finitely many types of charged matter particles in a candidate physical universe. Then $G$ must be one of the followings: 
$$\widetilde{\mr {SO}}(2n+1, 2)\,(n\ge 1),$$
and possibly $\widetilde{\mr {SL}}(2n, \bb R)\, (n\ge 3)$ and $\widetilde{\mr E}_{7(-25)}$. However, the charged matter particle count is $32$ when and only when $G=\widetilde{\mr {SO}}(11,2)$.

\vskip 10pt  The origin of mass, the question why seven spatial dimensions fail to expand to the macroscopic level or simply disappear,  and the question where the fundamental forces come from are all beyond us at the moment. Assume the starting point here is correct, then one can imagine that the thermal dynamics of the quantum states of fundamental matter particles should yield a clue to these questions. 

Even so, based on minimalism, naturality,  and simplicity, we can still make quiet a few further matches and predictions.

\section{Matching the gauge groups}\label{S: Further Matching}
In the real world universe in its current stage, four fundamental forces have been found: gravity and three gauge forces. The strength of these four forces goes like this: Strong force $>$ the electric force $>>$ the weak force $>>$ gravity. (Here $>$ means ``stronger than" and $>>$ means ``much stronger than".) Moreover, unlike other three forces, the weak force is a short range force \footnote{Strong force should be considered as a long range because gluons have rest mass zero. It only appears to be short range:  to separate quarks at one meter apart, one needs nearly infinity amount of energy.  Of course, the residual strong force between nucleons is short range, just like the residual electric force between atoms in a molecule. }. 

 At this moment, the best way of understanding these forces is via symmetry, a great idea starting from A. Einstein for gravity and then H. Weyl for the electric force, further developed for the strong force and the weak force from 1950s through early 1970s.  In this perspective, long range forces correspond to exact symmetry of the nature: gravity -- exact $\mr{SO}_0(3,1)$ symmetry \footnote{Here $\mr{SO}_0(3,1)$ denotes the identity component of $\mr{SO}(3,1)$}, the electric force -- exact $\mr{U}(1)_e$ symmetry, the strong force -- exact $\mr{SU}(3)_c$ symmetry. The weak force is obtained from a spontaneous symmetry breaking process: $\mr{U}(1)_Y\times \mr {SU}(2)_I$ is broken down to its diagonal $\mr{U}(1)_e$ Lie subgroup. For this reason, $\mr{U}(1)_Y\times \mr {SU}(2)_I$ is called a broken symmetry group.  
 
Note that, the exactness of the $\mr{SU}(3)_c$ symmetry explains why the up quarks of different colors have identical mass, and the broken electric-weak symmetry is the reason for the mass difference between up and down quarks. 
 
The appearance of these symmetries is quite mysterious, and that of symmetry broken is perhaps even more mysterious. However, these mysteries shall all be cleared up soon: these symmetries are all the symmetries of the set of visible states of the unique charged particle, moreover, such a symmetry is exact when it is also a symmetry of the set of all quantum states, and is broken otherwise. 

We have found some matches of the abstract universe $G=\widetilde{\mr {SO}}(11,2)$ with the real world universe. To further match the abstract universe $G$ with the real world universe,  the following facts shall be exploited hereon:
\begin{Fact} For the real world universe in its current stage,
\begin{itemize}
\item $\mr{U}(1)_e$, $\mr{SU}(3)_c$, $\mr{SO}_0(3,1)$ are exact  symmetry. 

\item $\mr{U}(1)_Y\times \mr{SU}(2)_I$ is a broken  symmetry. 

\end{itemize}

\end{Fact}
The actual match goes like this. First of all, since $G$ is big enough to contain
$$
G_{exact}:=\mr{U}(1)_e\times \mr{SU}(3)_c\times \mr{SO}_0(3,1)
$$ as a Lie subgroup (up to covering), $G_{exact}$ is indeed the exact symmetry. Secondly, we shall show in section \ref{S: visible states} that $V_{visible}$ is a module of
$$
G_{non-exact}:=G_{SM}\times \widetilde{\mr {SO}}(2,1),
$$
so, in view of the fact that $G_{non-exact}$ is not a Lie subgroup of
$G$ even up to covering, $G_{SM}$ must be broken. However, it is not clear at this moment why nature choses to break $G_{SM}$ down to $\mr{U}(1)_e\times \mr{SU}(3)_c$.

Based on the above analysis, up to covering, the abstract physical universe is provisionally taken as the one such that the unbroken gauge group $G_c$ is  $\mr{U}(1)_e\times \mr{SU}(3)_c$, and the unbroken space-time symmetry group $G_n$ is $\mr{SO}_0(3,1)$. 

It is not hard to see that $G_c$ can be as large as $\mr{Spin}(8)$.  However, since electron and down quarks have different masses,
we conclude that $G_c= \mr{U}(1)_e\times \mr{SU}(3)_c$ up to covering.
In principle, $G_n$ could be $\widetilde{\mr {SO}}(3,2)$, but we shall argue in the end of subsection \ref{SS: symmetry} that $G_n=\widetilde{\mr {SO}}(3,1)$. 

\begin{Thm}
An abstract physical universe that can possibly matches the real world universe at its current stage must be an one with $G=\widetilde{\mr {SO}}(11,2)$, unbroken gauge group $G_c=\mr{U}(1)_e\times_{\bb Z_3} \mr{SU}(3)_c$, and unbroken space-time symmetry group $G_n=\widetilde{\mr {SO}}(3,1)$.
\end{Thm}
From now on we shall always assume that $G_c=\mr{U}(1)_e\times_{\bb Z_3} \mr{SU}(3)_c$ and $G_n=\widetilde{\mr {SO}}(3,1)$. Note that $G_c$ is not a Lie subgroup of $\mr{Spin}(7)$, and
$$G_c=\mr{SU}(4)\cap (\mr{Spin}(2)\times_{\bb Z_2} \mr{Spin}(6))$$
as a Lie subgroup of $\mr{Spin}(8)$.

Similar analysis yields the following table: Let $D_{space}$ be the  the macroscopic spatial dimension, then
\begin{table}[H]
    \begin{tabular}{|c| c|}
        \hline
        $G_c$ contains & implies that $D_{space}$ must be  \\ \hline
       $\mr{SU}(3)_c$ & $\le 5$  \\ 
        \hline
 $\mr{U}(1)_e\times \mr{SU}(3)_c$ &  $\le 3$  \\ 
        \hline
   $G_{SM}$ &  $\le 1$  \\ 
        \hline
    \end{tabular}
\end{table}
\vskip -10 pt
\noindent It is now clear that the electric-weak force symmetry broken is due to a single dimension inequality:  $D_{space} > 1$. On the other hand, the fact that the electric and color gauge force symmetries is allowed to be unbroken is due to another dimension inequality: $D_{space}\le 3$. In fact, one can show that $3$ is the \emph{only} macroscopic spatial dimension that allows $G_{SM}$ to break down to ${\mr U}(1)_e\times \mr{SU}(3)_c$, but not to a bigger group.  

The existence of an abstract physical universe that matches all the facts on gauge symmetries as listed in the middle of this section is really something because, unless we are on the right track, these facts together might be strong enough to exclude all abstract physical universes.

\section{The visible states }\label{S: visible states}
To make a further match with the real world universe, one needs to understand the visible states. Assume for the moment that $G$ is broken to $$G':=\mr{Spin}(8)\times_{ \bb Z_2} \widetilde{\mr {SO}}(3,2)$$
and the macroscopic space dimension is $3$. What could be an visible state? A necessary condition is that the collection of visible states should be invariant under $G'$ and has GK dimension $3$. Another condition is that the ground states should be visible and the visible states should have significantly less energy. 

Let $V$ be the unitary lowest weight $(\frk{g}, K)$-module with the lowest weight
$$
(-{1\over 2}, -{1\over 2}, -{1\over 2}, -{1\over 2}, -{1\over 2},  5),
$$
i.e., the representation that corresponds to the charged matter particle in the abstract universe $G=\widetilde{\mr {SO}}(11,2)$. Under $\mr{Spin}(11)$ we have the decomposition
$$
V=\bigoplus_{I\ge 0} D_{\mr{Spin(11)}}(I),
$$ where $D_{\mr{Spin(11)}}(I)$ is the unitary highest weight module of $\mr{Spin}(11)$ with highest weight
$$
(I+{1\over 2}, {1\over 2}, {1\over 2}, {1\over 2},{1\over 2} ).
$$
When the universe cooled down to the stage for which $G$ is broken to $G'$, $D_{\mr{Spin(11)}}(I)$ breaks up into 
$\mr{Spin}(8)\times_{{\bb Z}_2} \mr{Spin}(3)$ - invariant pieces of the form
$$
D_{Spin(8)}^I(l)\otimes D^I_{Spin(3)}(m)\mbox{ or its C.C.},\; l,m \ge 0, l+m \le I. 
$$
Here, C.C. means complex conjugate, $D^I_{Spin(8)}(l)$ means the unitary highest weight module of $\mr{Spin}(8)$ with highest weight $
(l+{1\over 2}, {1\over 2}, {1\over 2},{1\over 2} )$,
and $D^I_{Spin(3)}(m)$ means the unitary highest weight module of $\mr{Spin}(3)$ with highest weight $m+{1\over 2}$. The extra index $I$ indicates that the invariant piece is inside $D_{\mr{Spin(11)}}(I)$. It is expected that, in any case I assume that, at this stage of our universe, the energy for states inside each of the above invariant piece is significantly high if $l>0$, therefore, the collection of visible states $V_{visible}$ is
$$\bigoplus_{I\ge 0}\bigoplus_{m=0}^I D_{Spin(8)}^I(0)\otimes D^I_{Spin(3)}(m) \bigoplus C.C=\bigoplus_{k\ge 0}V[k]$$
where
$$V[k]=\bigoplus_{m\ge 0} D_{Spin(8)}^{m+k}(0)\otimes D^{m+k}_{Spin(3)}(m) \bigoplus C.C.$$

One can verify that each $V[k]$ can be promoted to a unitary module of $G'$, i.e., 
$$S_+\otimes S_{\widetilde{\mr {SO}}(3,2)}[k]\bigoplus C.C.,$$
here  $S_+$ is the positive spin representation of $\mr{Spin}(8)$ and 
$S_{\widetilde{\mr {SO}}(3,2)}[k]$ is the unitary lowest weight representation of $\widetilde{\mr {SO}}(3,2)$ with lowest weight $(-{1\over 2}, 1)$, but with its grading degree shifted by $k$ so that its $n$-th component has degree $(n+k)$. In particular, this says that the particle content of these $V[k]$\,s are all the same:
a spin $1/2$ particle with $\mr{Spin}(8)$-charge $S_+$ and its antiparticle. Therefore, for $n\ge 1$, $V[n-1]$ is the $n$-th generation. Since each $V[k]$ has GK dimension two, $V_{visible}$ has GK dimension $3$. In short, we have proved that,
\vskip 10pt
{\em $D_{space} =3 \implies $  there are infinitely many generations of charged matter particles}.
\vskip 10pt
When $G$ is further broken to
$G_c\times _{\bb Z_2}G_n$, each $V[k]$ has exactly the particle content of a generation of quarks and leptons. This explains why there are at least three generations of quarks and leptons. On the other hand, there cannot be only three generations of quarks and leptons, otherwise, the macroscopic spatial dimension would be equal to two and an unnatural truncation would be chosen in
the direct sum $\bigoplus_{k\ge 0}V[k]$.  

It should be mentioned that $V[k]$ is a module of $G_{non-exact}$ (i.e., $G_{SM}\times\widetilde{\mr {SO}} (2,1)$):
$$
V[k]\cong (F\oplus F^*)\otimes \left( \bigoplus_{i\ge 0} S_{\widetilde{\mr {SO}} (2,1)}[i]\right),
$$
here $F$ is the Standard Model particle content introduced in section \ref{S: Postulate}, and $S_{\widetilde{\mr {SO}} (2,1)}[i]$ is the unitary lowest weight representation of $\widetilde{\mr {SO}}(2,1)$ with lowest weight ${1\over 2}$, but with its grading degree shifted by $i$. In particular, this proves that $V_{visible}$ is a module of $G_{SM}\times\widetilde{\mr {SO}} (2,1)$, a fact that has been used in section \ref{S: Further Matching}.

It is then clear that, the existing four fundamental forces all appear
in our theoretical framework, moreover, the symmetry responsible for the existence of these forces always transforms the quantum states of a quarks/leptons within a generation, i.e., leaves each generation invariant. Combining with the analysis in section \ref{S: Further Matching}, the electric-weak force symmetry group $\mr{U}(1)\times \mr{SU}(2)$, a Lie subgroup of $\mr{Spin}(4)$, must break down to $$\left(\mr{U}(1)\times \mr{SU}(2)\right)\cap \mr{Spin}(3),\; \mbox{i.e.},\; \mr{U}(1)_e,$$
moreover, this symmetry group transforms  a quark (or a lepton) within its generation, a prediction which is in agreement with the classical Standard Model. Therefore, this prediction contradicts to the assumption that the electric-weak symmetry transforms quarks (but not leptons) from different generations, the cornerstone which the Cabibbo-Kobayashi-Maskawa (CKM) theory \cite{Cabibbo63} is built upon.

A similar analysis concludes that there are infinitely many generations of type I dark matter particle, which are all CP neutral.

\section{Fifth fundamental force and Direct CP violation}\label{S: 5th force}
It is interesting to note that $V_{visible}$ can be {\em abstractly} promoted to the following $\mr{Spin}(8)\times \widetilde{\mr {SO}}(4,2)$ unitary module \footnote{This is implied by a branching rule for $(\widetilde{\mr{SO}}(4,2), \widetilde{\mr{SO}}(3,2))$.  }:
$$S_+\otimes S_{\widetilde{\mr {SO}}(4,2)}^+\bigoplus C.C.,$$
where $S_{\widetilde{\mr {SO}}(4,2)}^+$ is the unitary lowest weight representation of $\widetilde{\mr {SO}}(4,2)$ with lowest weight $(-{1\over 2}, -{1\over 2}, {3\over 2})$. Since $$G_c=\mr{U}(1)_e\times_{{\bb Z}_3}\mr{SU}(3)_c$$ is inside $\mr{Spin}(8)$, $V_{visible}$ is a $G_c\times \widetilde{\mr {SO}}(4,2)$ module, too. In view of the fact that $G_c$ cannot be inside 
$\mr{Spin}(7)$, this new $\widetilde{\mr {SO}}(4,2)$ symmetry must be broken for the same reason that the weak force symmetry must be broken: $\widetilde{\mr {SO}}(11,2)$ is not big enough to contain $G_c\times \widetilde{\mr {SO}}(4,2)$ up to covering. Indeed, this
$\widetilde{\mr {SO}}(4,2)$ symmetry is broken, otherwise, electron and muon would have the same amount of mass. In summary, the linear span of the quantum states of a fundamental matter particle (with both chiralities included) such as a quark and that of all its heavy analogues provides a {\em broken} unitary lowest weight representation for $\widetilde{\mr {SO}}(4,2)$ with lowest weight $(-{1\over 2}, -{1\over 2}, {3\over 2})$. 

Recall that, the breaking of the electric-weak symmetry to $\mr U(1)_e$ yields the long range electric force plus the short range weak force. Therefore, mathematically it is very convincing that the breaking of the conformal symmetry to $\widetilde{\mr {SO}}(3,1)$ yields the long range gravity force plus a new short range force. Just as weak force is much weaker than the electric force, this new force is expected to be much weaker than gravity, a main reason why it has not been experimentally observed yet. 

This new force {\em directly} violates the CP-symmetry: under CP, the $\widetilde{\mr {SO}}(4,2)$-module $S_{\widetilde{\mr {SO}}(4,2)}^+$ is turned into $S_{\widetilde{\mr {SO}}(4,2)}^-$ --- the unitary lowest weight representation of $\widetilde{\mr {SO}}(4,2)$ with lowest weight $({1\over 2}, -{1\over 2}, {3\over 2})$, a different module of $\widetilde{\mr {SO}}(4,2)$. In contrast, gravity does not violate the CP symmetry because $S_{\widetilde{\mr {SO}}(4,2)}^+$ and $S_{\widetilde{\mr {SO}}(4,2)}^-$ are the same when viewed as modules of $\widetilde{\mr {SO}}(3,2)$. 

\subsection{The Gravity-5th Symmetry Group}\label{SS: symmetry}
Although $V_{visible}$ is invariant under (the action of) $\mr{Spin}(6)\times \mr{Spin}(4)$, for the moment, there is no sure reason why
the actual strong [force] symmetry group is not $\mr{Spin}(6)$ rather than its subgroup $\mr{SU}(3)$ and why the electric-weak symmetry group is not $\mr{Spin}(4)$ ($=\mr{SU}(2)\times\mr{SU}(2)$) rather than its subgroup $\mr{U}(1)\times \mr{SU}(2)$. 

However, if the electric-weak force breaks parity symmetry, its symmetry group must be a Lie subgroup of $\mr{Spin}(4)$ in the following list:
$$
\mr{Spin}(4),\; \mr{U}(1)\times \mr{SU}(2),\; \mr{SU}(2)\times \mr{U}(1),\; \mr{U}(1)\times \mr{U}(1)
$$
from which nature picks up the middle one. Note that, since $\mr{Spin}(3)$ is the parity invariant Lie subgroup $\mr{Spin}(4)$, $\mr{U}(1)\times \mr{SU}(2)$ breaks down to its parity invariant Lie subgroup $(\mr{U}(1)\times \mr{SU}(2))\cap \mr{Spin}(3)$, i.e., $\mr{U}(1)_e$.

Since $V_{visible}$ is invariant under $\widetilde{\mr {SO}}(4,2)$, if gravity-5th force breaks CP symmetry, its symmetry group must be one of the followings:
$$
\mr {SO}(4,2),\; \mr {SO}(4,1),\;\mr {SO}(3,1)\times \mr{SO}(1,1)
$$ up to covering and connectedness. Therefore, if the electric-weak force can serve as a guidance, one can conclude that the gravity-5th symmetry group is  the middle group $\mr {SO}(4,1)$, i.e., $\widetilde{\mr {SO}}(4,1)$ more precisely. (By the way, $S_{\widetilde{\mr {SO}}(4,2)}^\pm$ are still irreducible when viewed as
$\widetilde{\mr {SO}}(4,1)$ modules.) However, $\mr {SO}(3,1)\times \mr{SO}(1,1)$ is also an attractive choice. Anyhow, a further investigation is needed in order to pin down 
the gravity-5th symmetry group $G''_n$ exactly. In any case, since $\mr{SO}(3,2)$ is the CP invariant Lie subgroup $\mr{SO}(4,2)$, $G_n''$ breaks down to its CP invariant Lie subgroup $G_n :=G_n''\cap \mr{SO}(3,2)$, i.e., $\mr{SO}(3,1)$, or $\widetilde{\mr {SO}}(3,1)$ more precisely. 

\subsection{Experimental Evidence for this 5th Force}
A salient feature of this 5th force is that it violates the CP symmetry and transforms a quark (or lepton) among its different generations. That leads to an obvious further prediction: decay of quarks from a higher generation to a lower generation and its related CP violation, a phenomena already observed in experiments by J. Cronin and V. Fitch about 50 years ago.

To explain this phenomena, the CKM theory was invented about 40 years ago. In the CKM theory, a key assumption is that the electric-weak symmetry mixes quarks (but not leptons) from different generations, something mysterious and not supported in our theoretical framework. So now there are two conflicting theories, the CKM theory and the one based on this 5th force. By testing them against the recent experimental data, one found that the CKM theory failed, but the one based on this 5th force withstood the test well, in both accuracy and precision.  For example, for the CKM matrix entry $V_{cd}$, we have \cite{mengmay12}
\begin{table}[H]
    \begin{tabular}{|c|c|c|c|}
        \hline
       & $\begin{matrix}\mbox{5th force based}\\ \mbox{computation}\end{matrix}$ & { experiment} & $\begin{matrix}\mbox{CKM theory based}\\ \mbox{computation}\end{matrix}$ \\ \hline
               $|V_{cd}|$ & $0.226\pm 0.034$ &  $0.230\pm 0.011$ &   $0.46\pm 0.48$ \\ 
        \hline
    \end{tabular}
\end{table}
\vskip -15pt
\noindent Consequently there is experimental evidence for the existence of this 5th force,  hence for the validity of the starting point of this paper.

The CP violation would provide a reasonable explanation for the matter-antimatter asymmetry in the real world universe in this stage; however, theoretical calculation based on the CKM theory as well as experimental measurement shows an excess far too small to account for the observed degree of asymmetry. This is another compelling piece of evidence for the existence of this 5th force and against the CKM theory.

\subsection{A Further Remark on this 5th Force}
Unlike the other fundamental forces, this 5th force owes its existence to the infinitely many generations of quarks and leptons. Also, in view of the CPT theorem, this 5th force is the only fundamental force that can distinguish the orientation of time in our universe. 

Two additional points are worth to mention. First, while the weak force can distinguish left from right, left-handed and right-handed electrons have the same mass; likewise, while this 5th force can distinguish matter from antimatter, electrons and positrons also have the same mass. Secondly, since the breaking of the electric-weak symmetry leads to the mass difference between electron and electron neutrino, the breaking of the gravity-5th symmetry leads to the mass difference between electrons, muons, and tauons. To put it differently, the breaking of the gravity-5th symmetry is responsible for the mass difference among different generations of the same particle. 

\subsection{The Role of this 5th Force in Our Universe}
In our life-supporting universe, the strong force is needed to form nuclei, the electric force is needed to form atoms and molecules, and gravity is needed to form galaxies, stars, and planets. These are long-range forces that can bind matter together. On the other hand, the weak force and this 5th force are short-range forces that can cause decay. 

One can speculate that this 5th force is responsible for the current accelerated expansion of our universe. Shortly after the Big Bang, almost all the particles were in high generation forms. Then, this 5th force together with the other fundamental forces caused the charged particles to decay into their 1st generation forms rapidly, releasing huge amounts of energy in a very short period of time. This might explain the occurrence of the cosmic inflation. Currently a lot of type I dark matter particles remain in their high generation forms and slowly decay to their 1st generation forms by this 5th force alone, releasing energy, albeit slowly. Although this current stage of energy release is much milder than that which occurred during the cosmic inflation, it is still sufficient to cause the accelerated expansion of our universe at the current stage. Interested readers may compare this speculation with that of Paul Steinhardt et al. \cite{Steinhardt}. 

While the charged matter particles are not CP neutral, type I dark matter particles are. Due to the CP violation of this 5th force as well as the charged matter and antimatter annihilation, dark matter particles, though very light, are much more abundant than the non-dark matter particles. This might be the reason why dark matter is the dominant type of matter in our universe.

\section{Some Loose Ends}\label{S: loose ends}
A lot has been deduced so far, but there are still old questions that have yet to be answered. For example, the origin of mass remains a mystery. Likewise, although symmetry has been used to obtain information about the fundamental forces, the origin of these forces is still not known (though one can imagine that all fundamental forces are just statistical phenomena involving the quantum states of elementary matter particles). Additionally, the theory proposed here also poses new problems, such as the role of type II dark matter in our universe. 

In principle $G_{exact}$ could be as large as
$G_{exact}':=\mr{Spin}(6)\times \mr{U}(1) \times \mr{SO}_0(3,2)$, but in fact it is actually this Lie subgroup of $G_{exact}'$:
$$
\mr{SU}(3)\times \mr{U}(1) \times \mr{SO}_0(3,1).
$$
Although it is not known why this is true, one thing is clear: $\mr{Spin}(6)/\mr{SU}(3)$ is compact with dimension $7$ and $\mr{SO}(3,2)/\mr{SO}(3,1)$ is non-compact with dimension $4$. This seems to suggest that at the moment our universe is anti de Sitter and our space-time has additional 7 compact spatial dimensions. If this speculation is true, it suggests that our universe will eventually contract because the energy released from the decay of type I dark matter in higher generation forms will eventually diminish, and that in turn implies an oscillating universe.

\section{Conclusions}\label{S: conclusion}
Fundamental physics today is based on two pillars:
\begin{itemize}
\item {\em quantum theory},
\item {\em Lorentz structure} (of 4D space-time). 
\end{itemize}
With that, one can build up suitable quantum field theory models to describe nature at the subatomic level.  However, one can not predict the existence of quarks and leptons from these two pillars, just as one cannot predict the spin of electron from the Schr\"odinger equation. In view of the fact that when the Schr\"odinger equation is replaced by the more refined Dirac equation, one can see not only the spin of electron, but also the positron, one might wonder whether one can see quarks, leptons and other new observations when Lorentz structure is replaced by some more refined structure.
Indeed this is the case for which Euclidean Jordan algebra is the more refined structure. In fact, with the replacement of Lorentz structure by Euclidean Jordan algebra, one can see a universe consisting of fundamental matter particles, both dark and non-dark particles. 

In case the Euclidean Jordan algebra is the one with conformal group $ \widetilde{\mr {SO}}(11,2)$, the abstract universe can have a fairly cold stage for which firstly, the exact symmetry group (up to covering) is $$\mr{U}(1)_e\times \mr{SU}(3)_c\times \widetilde{\mr {SO}}(3,1)$$ and secondly, the macroscopic space-time has 4D Lorentz structure. At this fairly cold stage, the unique non-dark particle becomes the quarks or leptons in the first generation plus their infinitely many heavier analogues; moreover, one can see that there are not only four existing fundamental forces but also a new 5th fundamental force.

The simple theory, which is based on replacing Lorentz structure by Euclidean Jordan algebra while keeping quantum theory, enables us to see many items in physics, such as quarks and leptons, and the existing four fundamental forces. It also predicts many items, such as dark matter and the 5th force. Finally, it solves a few mysteries on the conceptual level, such as the quark generation mystery and the matter/anti-matter asymmetry.  There are quite a few shocking mismatches with current opinions, scattered around in various places of this article. However, as far as matching hard facts at the discrete quantitative level is concerned, no mismatch is found so far, in fact, the theory shows a remarkable internal consistency. Moreover, the recent experimental measurements of the CKM matrix entries provide a compelling piece of quantitative evidence for the existence of the predicted 5th force. All of these indicate that this theory is probably on the right track.  

Finally, it should be pointed out that, assuming the proposed modification of twin pillars is correct, the conclusions derived here can be trusted because they are obtained by employing the most reliable tool in physics, i.e. symmetry analysis. 

\section{Acknowledgement}

I would like to thank R. Howe for his enthusiasm on this piece of work when it just got started. I would like to thank C. Taubes for a careful reading of the 1st draft, for his advices, comments, and knowledges about the Standard Model of particle physics. I would like to thank Sir M. Atiyah for helping me to communicate this article to one of his  colleagues in physics. I would like to thank all three of them for support and encouragement all along. Finally, I would like to thank J. Baez for his valuable advices on the presentation of this article.

\end{document}